 \documentclass[showpacs,twocolumn,preprintnumbers,amsmath,amssymb]{revtex4}
 \usepackage{dcolumn}
 \usepackage{bm}

\begin{document}

\preprint{}

\title{Knotted Topological Phase Singularities of Electromagnetic Field}

\author{Ji-Rong Ren }
\author {Tao Zhu }
\thanks{Corresponding author. Email : zhut05@lzu.cn }
\author {Shufan Mo}
\affiliation{Institute of Theoretical Physics, Lanzhou University,
Lanzhou 730000, P. R. China}

\date{\today}

\begin{abstract}
In this paper, knotted objects (RS vortices) in the theory of
topological phase singularity in electromagnetic field have been
investigated in details. By using the $\phi$-mapping topological
current theory proposed by \emph{Prof. Duan}, we rewrite the
topological current form of RS vortices and use this topological
current we reveal that the Hopf invariant of RS vortices is just the
sum of the linking and self-linking numbers of the knotted RS
vortices. Furthermore, the conservation of the Hopf invariant in the
splitting, the mergence and the intersection processes of knotted RS
vortices is also discussed.
\end{abstract}



 \maketitle

\section{Introduction}
Topological phase singularities as topological objects of wave
fields appear in a variety of physical, chemical, and biological
scenarios, such as the quantized vortices in superfluid or
superconductor systems\cite{supflu,quanvortex}, the vortices phase
singularities in Bose-Einstein condensates\cite{BEC}, the
streamlines or singularities in quantum mechanical
wave-functions\cite{quanwave, quanvor}, the optical vortices in
optical wave systems\cite{opt1, opt2, opt3, optic, op-top-zhu,
zhu2}, and the vortex filaments in chemical reaction and molecular
diffusion\cite{chemwave, fil-zhu}.

In particular, the study of phase singularities in electromagnetic
field or in optics has evolved into a separate area of research,
both theoretical and experimental, called singular
optics\cite{sinopt}. Various types of phase singularities in optics
have been found. In the usual scalar theory of light in optics, when
the vector nature of light can be neglected, one may build a single
complex scalar field to describe the light. The optical vortices
emerge as phase singularities of the complex scalar field, which are
located at the zeros of the field. But for a full electromagnetic
field which can be described by a complex vector, the phase
singularity is very difficult to define because the requirement that
all the field components can not simultaneously vanish. However, it
is interesting from a fundamental theoretical viewpoint to
investigate the phase singularities of the full electromagnetic
field. Such singularities are just the Riemann-Silberstein (RS)
vortices which associated with the zeros of square of RS vector, and
was firstly studied recently by I. Bialynicki-Birula and Z.
Bialynicka-Birula\cite{rs1}. Since the first proposal of RS
vortices, a great many of works have been focused on the basic
character of this phase singularities, especially the geometrical
and topological character\cite{rs2, rs3, rs3-topo}. In the
Ref.\cite{rs3-topo}, the authors have derived the topological
current structures of the RS vortex line.

On the other hand, the phase singularities usually form a net of
lines in three dimensional space, and a very important case is that
they are closed and knotted curves. The knot-like configurations
exist in a variety of physical scenarios, including Bose-Einstein
condensations\cite{BEC1}, chemical reaction and molecular diffusion
systems\cite{chemwave, fil-zhu}, optical wave systems\cite{opt3,
op-top-zhu, zhu2, opk1} and field theory\cite{fie, fie2}. For a
knotted family, it is well known that there are important
characteristic numbers to describe its topology, such as the
self-linking and the linking numbers. So in research into knotted
configurations in physics, one should pay much attention to these
knotted characteristics. In this paper, we will use the topological
viewpoint to study the knotted topological phase singularities. i.e.
the RS vortices with the Hopf invariant which usually can be used to
describe the linkage of the knotted family in
mathematics\cite{Hopf}, and reveal the inner relationship between
the Hopf invariant and the topological knotted characteristic
numbers of knotted RS vortices. Furthermore, the conservation of the
Hopf invariant in the splitting, the mergence and the intersection
processes is also discussed in details.

\section{A brief review of RS vortices and their topological structures}
In the theory of the complex form of the Maxwell equations, the
electric and magnetic field vectors can be replaced by a complex
vector $\vec{F}$ which called RS vector\cite{rs1, rs2},
\begin{eqnarray}
\vec{F}=\frac{1}{\sqrt{2}}(\vec{E}+i\vec{B}).
\end{eqnarray}
The Maxwell equations in free space written in terms of $\vec{F}$
read:
\begin{eqnarray}
&&i \partial_t \vec{F}=\nabla \times \bar{F},\nonumber\\
&& \bigtriangledown \cdot \vec{F}=0.
\end{eqnarray}
The RS vector offers a very convenient representation of the
electromagnetic field, especially in the study of phase
singularities. Since $\vec{F}^2$ is a complex sum of two
electromagnetic invariants, we define the phase of the
electromagnetic filed $\varphi (x)$ as half of the phase of square
of the RS vector
\begin{equation}
\vec{F}^2(x)=e^{2i\varphi(x)}\|\vec{F}^2(x) \|.
\end{equation}
In full analogy with equation of nonrelativistic wave mechanics, we
may define a "velocity" four-vector $v_\mu$ is
\begin{equation}
v_\mu=\frac{(\vec{F}^2)^*\partial_\mu
\vec{F}^2-\vec{F}^2\partial_\mu(\vec{F}^2)^*}{4i\|\vec{F}^2\|^2}.
\end{equation}
Because $\vec{F}^2$ is a complex sum of two electromagnetic
invariants, it can also be written as
\begin{equation}
\vec{F}^2=S+iP=\frac{1}{2}(\vec{E}^2-\vec{B}^2)+i\vec{E}\cdot
\vec{B},
\end{equation}
then the velocity $v_\mu$ is true relativistic four-vector
\begin{equation}
v_\mu=\frac{S\partial_\mu P-P\partial_\mu S}{2(S^2+P^2)},
\end{equation}
in three-dimensional space, the spatial component of the
four-velocity is
\begin{equation}
\vec{v}=\frac{S\nabla P-P\nabla S}{2(S^2+P^2)}.\label{3-vector}
\end{equation}

For convenience, we write $\vec{F^2}$ as $\psi$, that is to say
\begin{eqnarray}
\vec{F^2}=\psi(\vec{x},t)=\phi^1(\vec{x},t)+i\phi^2(\vec{x},t),
\end{eqnarray}
where $\phi^1=S$ and $\phi^2=P$. The velocity Eq.(\ref{3-vector})
can be rewritten as
\begin{equation}
\vec{v}=\frac{1}{4i}\frac{1}{\|\psi\|^2}(\psi^*\nabla\psi-\nabla\psi^*\psi)=\nabla\varphi,\label{velocity}
\end{equation}
it becomes the gradient of the phase factor $\varphi$. Owing to the
works done by Bialynicki-Birula and Bialynicka-Birula, the phase
singularities of the electromagnetic field can be well defined by
using of $\varphi$, which is just the phase factor of the
electromagnetic field. This singularities are the so-called RS
vortex lines which can be found at the intersection of the $S=0$ and
$P=0$ surfaces.

Using the definition of the RS vortices, it is interesting to study
the topological properties of these topological phase singularities.
In Ref.\cite{rs3-topo}, by making use of the topological viewpoint
the authors have obtained the topological inner structures of the RS
vortices, and they have found that the RS vortices are classified by
Hopf index, Brouwer degree in geometry. In the following discussions
of this section, for convenience to discuss the knotted topology of
RS vortices, we will rewrite the topological form of the RS vortices
which have been derived in Ref.\cite{rs3-topo}.

According to the $\phi$-mapping theory proposed by Prof.
Duan\cite{fie2,tc}, the topological density vector current
$\vec{\Omega}$ can be defined as
\begin{eqnarray}
\vec{\Omega}=\frac{1}{\pi}\nabla\times\vec{v},\label{tpc}
\end{eqnarray}
Form Eq.(\ref{tpc}), we directly obtain a trivial curl-free result:
$\vec{\Omega}=\frac{1}{\pi}\nabla\times\nabla\varphi=0$. But in
topology, because of the existence of the phase singularities. i.e.
RS vortices, $\vec{\Omega}$ does not vanish\cite{quanvortex}. So in
the following discussions, we will study what the exact expression
for $\vec{\Omega}$ is in topology.

Introducing the unit vector $n^a=\phi^a/\|\phi\| (a=1, 2;
n^an^a=1)$, one can reexpress the velocity field $\vec{v}$ as
\begin{equation}
\vec{v}=\frac{1}{2}\epsilon_{ab}n^a\nabla n^b,
\end{equation}
and $\vec{\Omega}$ is
\begin{equation}
\Omega^i=\frac{1}{2\pi}\epsilon^{ijk}\epsilon_{ab}\partial_jn^a\partial_kn^b,~~~~~i,j,k=
1, 2, 3.\label{topovorticity}
\end{equation}
Obviously, it is just the topological current of the RS vortices in
three dimensional space. It is known that in $\phi$-mapping theory,
the topological current $\vec{\Omega}$ is rewritten as
\begin{equation}
\Omega^i=\delta^2(\vec{\phi})D^i(\frac{\phi}{x}),\label{delat}
\end{equation}
where the $D^i(\frac{\phi}{x})$ is the vector Jacobians of
$\psi(\vec{r})$, and it is defined as
\begin{equation}
D^i(\frac{\phi}{x})=\frac{1}{2}\epsilon^{ijk}\epsilon_{ab}\partial_j\phi^a\partial_k\phi^b,
\end{equation}
we can see from the expression (\ref{delat}) that $\vec{\Omega}$ is
non-vanishing only if $\vec{\phi}=0$, i.e., the existence of the RS
vortices, so it is necessary to study these zero solutions of
$\vec{\phi}$. In three dimensions space, these solutions are some
isolate zero lines, which are the so-called RS vortices in three
dimensional space.

Under the regular condition $$D(\phi/x)\neq0,$$ the general
solutions of
\begin{equation}
\phi^1(t,x^1,x^2,x^3)=0,~~~~~ \phi^2(t,x^1,x^2,x^3)=0
\end{equation}
can be expressed as
\begin{equation}
x^1=x^1_k(s,t), ~~~x^2=x^2_k(s,t),~~~x^3=x^3_k(s,t),
\end{equation}
which represent $N$ isolated singular strings $L_k$ with string
parameter $s$ $(k=1,2,\cdots,N)$. These singular strings solutions
are just the RS vortices solutions in three dimensions space.

In $\delta$-function theory\cite{deltfun}, one can obtain in three
dimensions space
\begin{equation}
\delta^2(\vec{\phi})=\sum_{k=1}^N \beta_k
\int_{L_k}\frac{\delta^3(\vec{x}-\vec{x}_k(s))}{|D(\frac{\phi}{u})|_{\Sigma_k}}ds,\label{delta3}
\end{equation}
where
$$D(\frac{\phi}{u})|_{\Sigma_k}=\frac{1}{2} \epsilon^{jk}\epsilon_{mn}\frac{\partial\phi^m}{\partial
u^j}\frac{\partial\phi^n}{\partial u^k}, $$ and $\Sigma_k$ is the
$k$th planar element transverse to $L_k$ with local coordinates
$(u^1,u^2)$. The $\beta_k$ is the Hopf index of $\phi$ mapping,
which means that when $\vec{x}$ covers the neighborhood of the zero
point $\vec{x}_k(s)$ once, the vector field $\phi$ covers the
corresponding region in $\phi$ space $\beta_k$ times. Meanwhile the
direction vector of $L_k$ is given by
\begin{equation}
\frac{dx^i}{ds}|_{x_k}=\frac{D^i(\phi/x)}{D(\phi/u)}|_{x_k}.\label{direction}
\end{equation}
Then from Eq.(\ref{delta3}) and Eq.(\ref{direction}) one can obtain
the inner structure of $\Omega^i$:
\begin{equation}
\Omega^i=\sum_{k=1}^N W_k
\int_{L_k}\frac{dx^i}{ds}\delta^3(\vec{x}-\vec{x}_k(s))ds,\label{topo}
\end{equation}
where $W_k=\beta_k\eta_k$ is the winding number of $\vec{\phi}$
around $L_k$, with $\eta_k=sgn D(\phi/u)|_{\vec{x}_k}=\pm1$ being
the Brouwer degree of $\phi$ mapping. The sign of Brouwer degrees
are very important, the $\eta_k=+1$ corresponds to the vortex, and
$\eta_k=-1$ corresponds to the antivortex. The integer number $W_k$
measures windings of the phase around the phase singularities, and
is also called the topological charge  of the RS vortex.  Hence the
topological charge of the RS vortices $L_k$ is
\begin{equation}
Q_k=\int_{\Sigma_k}\Omega^id\sigma_i=W_k.\label{topochar}
\end{equation}
The Eq.(\ref{topochar}) shows us that the topological current
$\vec{\Omega}$ describes the density of RS vortices in space.
Therefore, we call the topological current $\vec{\Omega}$ the
topological charge current density  of the RS vortices.

The results in this section show us the topological inner structure
of the topological charge current density $\vec{\Omega}$. The RS
vortices are classified by Hopf indices and Brouwer degrees. But we
must note that these topological numbers describe the topological
properties only for RS vortices, not for the phase singularities of
the electromagnetic field. The definition of the topological numbers
of phase singularities are different with the RS vortices. Though
the sites of RS vortices are just the points of the phase
singularities, from the previous definition of the phase of the
electromagnetic field, we know that the phase of the electromagnetic
filed $\varphi (x)$ is half of the phase of square of the RS vector.
In other words, the phase of square of the RS vector is
$\chi=2\varphi$. In the above discussions, the winding number of the
$\phi$-mapping is relate to the phase $\chi$ .i.e.,
\begin{eqnarray}
W_k=\frac{1}{2\pi}\oint_l d\chi,
\end{eqnarray}
where the closed path $l$ surrounds the $k$th RS vortex. For the
phase singularities, the winding number of the singular point is
\begin{eqnarray}
W'_k=\frac{1}{2\pi}\oint_l d\varphi.
\end{eqnarray}
Here the winding number $W'_k$ measures the strength of the phase
singularities. It is easy to see that there are two different
expressions of the winding number relate to a same singular point,
but they are not independent with each other because
$\chi=2\varphi$. In our following discussions, we only use the
topological numbers which relate to RS vortices, and note that they
are also valid for the phase singularities of electromagnetic field.

\section{The Hopf Invariant of the Knotted RS vortices}

In this section, let us begin to discuss the topological properties
of knotted RS vortices. It is well known that the Hopf invariant is
an important topological invariant to describe the topological
characteristics of the knot family. In a closed three-manifold $M$,
the Hopf invariant is defined as\cite{fil-zhu, Hopf}
\begin{equation}
H=\frac{1}{2\pi}\int_M A_i \Omega^id^3x\label{held},
\end{equation}
in which $A_i$ is a "induced Abelian gauge potential" constructed
with the complex function $\vec{F}^2$, and
$$A_i=\epsilon_{ab}n^a\partial_in^b=2v_i=\partial_i\chi.$$

Substituting Eq. (\ref{topo}) into Eq. (\ref{held}), one can obtain
\begin{equation}
H=\frac{1}{2\pi}\sum_{k=1}^{N}W_k \int_{L_k}\vec{A}\cdot
d\vec{x}\label{helint1},
\end{equation}
for closed and knotted lines, i.e., a family of knots $\xi_k (k=1,
2, \ldots, N)$, Eq. (\ref{helint1}) becomes
\begin{equation}
H=\frac{1}{2\pi}\sum_{k=1}^{N}W_k \oint_{\xi_k}\vec{A}\cdot
d\vec{x}.\label{lak}
\end{equation}
This is a very important expression. Consider a transformation of
$\vec{F}^2$: $\psi'=e^{i\chi}\psi$, this gives the U(1) gauge
transformation of $\vec{A}$: $A'_i=A_i+\partial_i \chi$, where
$\chi$ is the phase factor denoting the U(1) gauge transformation.
It is seen that the $\partial_i \chi$ term in Eq. (\ref{lak})
contributes nothing to the integral $H$ when RS vortices are closed,
hence the expression (\ref{lak}) is invariant under the U(1) gauge
transformation.

It is well known that many important topological numbers are related
to a knot family such as the self-linking number and Gauss linking
number. In order to discuss these topological numbers of knotted RS
vortices, we define Gauss mapping:
\begin{equation}
\vec{m}: S^1 \times S^1 \rightarrow S^2,
\end{equation}
where $\vec{m}$ is a unit vector
\begin{equation}
\vec{m}(\vec{x},
\vec{y})=\frac{\vec{y}-\vec{x}}{|\vec{y}-\vec{x}|},\label{unit}
\end{equation}
where $\vec{x}$ and $\vec{y}$ are two points, respectively, on the
knots $\xi_k$ and $\xi_l$ (in particular, when $\vec{x}$ and
$\vec{y}$ are the same point on the same knot $\xi$ , $\vec{n}$ is
just the unit tangent vector $\vec{T}$ of $\xi$ at $\vec{x}$ ).
Therefore, when $\vec{x}$ and $\vec{y}$ , respectively, cover the
closed curves $\xi_k$ and $\xi_l$ once, $\vec{n}$ becomes the
section of sphere bundle $S^2$. So, on this $S^2$ we can define the
two-dimensional unit vector $\vec{e}=\vec{e}(\vec{x}, \vec{y})$.
$\vec{e}$, $\vec{m}$ are normal to each other, i.e. ,
\begin{eqnarray}
&&\vec{e}_1\cdot\vec{e}_2=\vec{e}_1\cdot\vec{m}=\vec{e}_2\cdot\vec{m}=0,
\nonumber\\&&\vec{e}_1\cdot\vec{e}_1=\vec{e}_2\cdot\vec{e}_2=\vec{m}\cdot\vec{m}=1.
\end{eqnarray}
In fact, the gauge potential $\vec{A}$ can be decomposed in terms of
this two-dimensional unit vector $\vec{e}$:
$A_i=\epsilon_{ab}e^a\partial_i e^b-\partial_i\chi$, where $\chi$ is
a phase factor\cite{quanvortex}. Since one can see from the
expression $\vec{\Omega}=\frac{1}{\pi}\nabla\times
\vec{v}=\frac{1}{2\pi}\nabla\times \vec{A}$ that the
$(\partial_i\chi)$ term does not contribute to the integral $H$,
$A_i$ can in fact be expressed as
\begin{equation}
A_i=\epsilon_{ab}e^a\partial_ie^b.
\end{equation}
Substituting it into Eq.(14), one can obtain
\begin{equation}
H=\frac{1}{2\pi}\sum_{k=1}^{N}W_k
\oint_{\xi_k}\epsilon_{ab}e^a(\vec{x},
\vec{y})\partial_ie^b(\vec{x}, \vec{y})dx^i.\label{hel}
\end{equation}
Noticing the symmetry between the points $\vec{x}$ and $\vec{y}$ in
Eq.(\ref{unit}), Eq.(\ref{hel}) should be reexpressed as
\begin{equation}
H=\frac{1}{2\pi}\sum_{k, l=1}^N W_k W_l \oint_{\xi_k}
\oint_{\xi_l}\epsilon_{ab}\partial_i e^a\partial_j e^bdx^i\wedge
dy^j.\label{hel2}
\end{equation}
In this expression there are three cases: (1) $\xi_k$ and $\xi_l$
are two different RS vortices $(\xi_k\neq\xi_l)$, and $\vec{x}$ and
$\vec{y}$ are therefore two different points $(\vec{x}\neq\vec{y})$;
(2) $\xi_k$ and $\xi_l$ are the same RS vortices $(\xi_k=\xi_l)$,
but $\vec{x}$ and $\vec{y}$ are two different points
$(\vec{x}\neq\vec{y})$; (3) $\xi_k$ and $\xi_l$ are the same RS
vortices $(\xi_k=\xi_l)$, and $\vec{x}$ and $\vec{y}$ are the same
points $(\vec{x}=\vec{y})$. Thus, Eq.(\ref{hel2}) can be written as
three terms:
\begin{eqnarray}
&&H=\sum_{k=1(k=l, \vec{x\neq}\vec{y})}^N \frac{1}{2\pi}W_k^2
\oint_{\xi_k} \oint_{\xi_k} \epsilon_{ab} \partial_i e^a\partial_j
e^b dx^i \wedge dy^j \nonumber\\&&+\frac{1}{2\pi}\sum_{k=1}^N W_k^2
\oint_{\xi_k} \epsilon_{ab} e^a\partial_i e^bdx^i
\nonumber\\&&+\sum_{k, l=1(k\neq l)}^N \frac{1}{2\pi}W_k W_l
\oint_{\xi_k} \oint_{\xi_l} \epsilon_{ab}
\partial_i e^a\partial_j e^b
dx^i \wedge dy^j.\label{hel3}
\end{eqnarray}
By making use of the relation
$\epsilon_{ab}\partial_ie^a\partial_je^b=\frac{1}{2}\vec{m}\cdot(\partial_i\vec{m}\times\partial_j\vec{m})$\cite{rela},
the Eq.(\ref{hel3}) is just
\begin{eqnarray}
H&=&\sum_{k=1(\vec{x}\neq\vec{y})}^N \frac{1}{4\pi}W_k^2
\oint_{\xi_k}\oint_{\xi_k} \vec{m}^*(dS) \nonumber \\ &&
+\frac{1}{2\pi}\sum_{k=1}^N W_k^2 \oint_{\xi_k} \epsilon_{ab}e^a
\partial_ie^bdx^i \nonumber \\ &&
+\sum_{k, l=1(k\neq l)}^N \frac{1}{4\pi} W_k W_l
\oint_{\xi_k}\oint_{\xi_l} \vec{m}^*(dS),\label{hel4}
\end{eqnarray}
where
$\vec{m}^*(dS)=\vec{m}\cdot(\partial_i\vec{m}\times\partial_j\vec{m})dx^i\wedge
dy^j(\vec{x}\neq\vec{y})$ denotes the pullback of the $S^2$ surface
element.

In the following we will investigate the three terms in the
Eq.(\ref{hel4}) in detail. Firstly, the first term of
Eq.(\ref{hel4}) is just related to the writhing number\cite{writnum}
$Wr(\xi_k)$ of $\xi_k$
\begin{equation}
Wr(\xi_k)=\frac{1}{4\pi}\oint_{\xi_k}\oint_{\xi_l}
\vec{m}^*(dS).\label{writhing}
\end{equation}
For the second term, one can prove that it is related to the
twisting number $Tw(\xi_k)$ of $\xi_k$
\begin{eqnarray}
\frac{1}{2\pi}\oint_{\xi_k}\epsilon_{ab}e^a\partial_ie^bdx^i
&&=\frac{1}{2\pi}\oint_{\xi_k}(\vec{T}\times\vec{V})\cdot
d\vec{V}\nonumber\\&&=Tw(\xi_k),\label{twisting}
\end{eqnarray}
where $\vec{T}$ is the unit tangent vector of knot $\xi_k$ at
$\vec{x}$ ($\vec{m}=\vec{T}$ when $\vec{x}=\vec{y}$) and $\vec{V}$
is defined as
$e^a=\epsilon^{ab}V^b(\vec{V}\perp\vec{T},\vec{e}=\vec{T}\times\vec{V})$.
In terms of the White formula\cite{writfor}
\begin{equation}
SL(\xi_k)=Wr(\xi_k)+Tw(\xi_k),\label{self}
\end{equation}
we see that the first and the second terms of Eq.(\ref{hel4}) just
compose the self-linking numbers of knots.

Secondly, for the third term, one can prove that
\begin{eqnarray}
&&\frac{1}{4\pi}\oint_{\xi_k}\oint_{\xi_l}
\vec{m}^*(dS)\nonumber\\&&=\frac{1}{4\pi}\epsilon^{ijk}\oint_{\xi_k}dx^i\oint_{\xi_l}dy^j
\frac{(x^k-y^k)}{\|\vec{x}-\vec{y}\|^3}\nonumber\\&&=Lk(\xi_k,\xi_l)~~(k\neq
l),\label{linking}
\end{eqnarray}
where $Lk(\xi_k,\xi_l)$ is the Gauss linking number between $\xi_k$
and $\xi_l$\cite{writnum}. Therefore, from Eqs.(\ref{writhing}),
(\ref{twisting}), (\ref{self}) and (\ref{linking}), we obtain the
important result:
\begin{equation}
H=\sum_{k=1}^N W_{k}^2 SL(\xi_{k})+\sum_{k,l=1(k\neq
l)}^NW_kW_lLk(\xi_k,\xi_l).
\end{equation}
This precise expression just reveals the relationship between $H$
and the self-linking and the linking numbers of the RS vortices
knots family\cite{writnum}. Since the self-linking and the linking
numbers are both the invariant characteristic numbers of the RS
vortices knots family in topology, $H$ is an important topological
invariant which required to describe the linked RS vortices.

\section{The conservation of Hopf Invariant}

In this section we will simply discuss the conservation of the Hopf
invariant in the branch processes of knotted RS vortices.

In the previous work done in Ref.\cite{rs3-topo}, authors have point
out that in the evolution of RS vortices, the splitting, the
mergence and the intersection processes may occur when the condition
$D(\phi/x)\neq0$ fails. In these branch processes, we note that the
sum of the topological charges of final RS vortices must be equal to
that of the initial vortices at the bifurcation point. This
conclusion is always valid because it is in topological level. So we
have,

(a) for the case that one vortex $L$ split into two vortices $L_{1}$
and $L_{2}$, we have $W_{L}=W_{L_1}+W_{L_2}$;

(b) two vortices $L_{1}$ and $L_{2}$ merge into one vortices:
$W_{L_1}+W_{L_2}=W_{L}$;

(c) two vortices $L_{1}$ and $L_{2}$ meet, then depart as other two
vortices $L_{3}$ and $L_{4}$: $W_{L_1}+W_{L_2}=W_{L_3}+W_{L_4}$.

In the following we will show that when the branch processes of
knotted RS vortices occur as above, the Hopf invariant is preserved:

(A) The splitting case. We consider one knot $\xi$ split into two
knots $\xi_{1}$ and $\xi_{2}$ which are of the same seif-linking
number as $\xi$ $(SL(\xi)=SL(\xi_{1})=SL(\xi_{2}))$. And then we
will compare the two number $H_\xi$ and $H_{\xi_{1}+\xi_{2}}$ (where
$H_{\xi}$ is the contribution of $\xi$ to $H$ before splitting, and
$H_{\xi_1+\xi_{2}}$ is the total contribution of $\xi_{1}$ and
$\xi_2$ to $H$ after splitting. First, from the above text we have
$W_{\xi}=W_{\xi_{1}}+W_{\xi_{2}}$ in the splitting process. Second,
on the one hand, noticing that in the neighborhood of bifurcation
point, $\xi_{1}$ and $\xi_{2}$ are infinitesimally displace from
each other; on the other hand, for a knot $\xi$ its self-linking
number $SL(\xi)$ is defined as $SL(\xi)=Lk(\xi,\xi_{V})$, where
$\xi_{V}$ is another knot obtained by infinitesimally displacing
$\xi$ in the normal direction $\vec{V}$\cite{writnum}. Therefore
$SL(\xi)=SL(\xi_{1})=SL(\xi_{2})=Lk(\xi_{1},\xi_{2})=Lk(\xi_{2},\xi_{1})$,
and $Lk(\xi,\xi'_{k})=Lk(\xi_{1},\xi'_{k})=Lk(\xi_{2},\xi'_{k})$
(where $\xi'_{k}$ denotes another arbitrary knot in the
family($\xi'_{k}\neq \xi, \xi'_{k}\neq \xi_{1,2}$)). Then, third, we
can compare $H_{\xi}$ and $H_{\xi_{1}}+_{\xi_{2}}$ before splitting,
\begin{eqnarray}
H_{\xi}=W^2_{\xi}SL(\xi)+\sum_{k=l(\xi'_{k}\neq
\xi)}^{N}2W_{\xi}W_{\xi'_{k}}Lk(\xi,\xi'_{k}),
\end{eqnarray}
where $Lk(\xi,\xi'_{k})=Lk(\xi'_{k},\xi)$; after splitting,
\begin{eqnarray}
H_{\xi_1+\xi_2}&=&W^{2}_{\xi_1}SL(\xi_1)+W^{2}_{\xi_{2}}SL(\xi_{2})+
2W_{\xi_{1}}W_{\xi_{2}}Lk(\xi_{1},\xi_{2})\nonumber\\
&+&\sum_{k=l(\xi'_k\neq
\xi_{1,2})}^{N}2W_{\xi_{1}}W_{\xi'_{k}}Lk(\xi_{1},\xi'_{k})\nonumber\\
&+&\sum_{k=l(\xi'_{k}\neq\xi_{1,2})}^{N}2W_{\xi_{2}}W_{\xi'_{k}}Lk(\xi_{2},\xi'_{k}).
\end{eqnarray}
Comparing (35) and (36), we have
\begin{eqnarray}
H_\xi=H_{\xi_1+\xi_2}
\end{eqnarray}
This means that in the splitting process the Hopf invariant is
conserved.

(B) The mergence case. We consider two knots $\xi_{1}$ and
$\xi_{2}$, which are of the same self-linking number, merge into one
knot $\xi$ which is of the same self-linking number as $\xi_{1}$ and
$\xi_{2}$. This is obviously the inverse process of the above
splitting case, therefore we have
\begin{eqnarray}
H_{\xi_1+\xi_2}=H_\xi.
\end{eqnarray}

(C) The intersection case. This  case is related to the collision of
two knots. we consider that two knots $\xi_{1}$ and $\xi_{2}$, which
are of the same self-linking number, meet, and then depart as other
two knots $\xi_{3}$ and $\xi_{4}$ which are of the same self-linking
number as $\xi_{1}$ and $\xi_{2}$. This process can be identified to
two sub-processes: $\xi_{1}$ and $\xi_{2}$ merge into one knot
$\xi$, and then $\xi$ split into $\xi_{3}$ and $\xi_{4}$. Therefore,
from the above two cases (B) and (A) we have
\begin{eqnarray}
H_{\xi_1+\xi_2}=H_{\xi_3+\xi_4}
\end{eqnarray}
Therefore we acquire the result that, in the branch processes during
the evolution of knotted RS vortices (splitting, mergence, and
intersection), the Hopf invariant is preserved.

\section{Conclusion}
In this paper, knotted objects in the theory of topological phase
singularity in electromagnetic field have been investigated in
details. By making use of the $\phi$-mapping topological current
theory, we rewrite the topological inner structure of the RS
vortices. Because the phase of the electromagnetic filed
$\varphi(x)$ is half of the phase of square of the RS vector, we
point out that the topological numbers for phase singularities are
different with the same numbers of RS vortices. So, for the same
singular point where the RS vortex and phase singularity sits, the
topological numbers have two definitions, one for RS vortices and
another for phase singularities. Because this two definitions are
not independent, in this paper we adopt the definition for RS
vortices.

Furthermore, we study the knot topology of knotted RS vortices in
terms of Hopf invariant. It is revealed that the Hopf invariant $H$
is just the total sum of all the self-linking and linking numbers of
knotted family. At last, it is shown that $H$ is preserved in the
splitting, the mergence and the intersection processes of knotted RS
vortices.

\begin{acknowledgments}
This work was supported by the National Natural Science Foundation
of China and the Cuiying Programme of Lanzhou University.

 \end{acknowledgments}

\end{document}